\documentclass[preprint,amsmath,amssymb,aps,physrev,superscriptaddress]{revtex4-2}

\usepackage{graphicx}
\usepackage{dcolumn}
\usepackage{bm}
\usepackage{hyperref}
\usepackage[mathlines]{lineno}
\usepackage{comment}

\usepackage{url}

\usepackage{longtable}

\usepackage{enumitem}
\setlist{nolistsep}

\usepackage{xcolor}
\definecolor{myOrange}{rgb}{1,0.5,0.}
\definecolor{myGreen}{rgb}{0.0,0.6,0.1}

\newcommand{\removetext}[1]{} 

\newcommand{\nineH}        {$\sqrt{s}~=~0.9$~Te\kern-.1emV\xspace}
\newcommand{\seven}        {$\sqrt{s}~=~7$~Te\kern-.1emV\xspace}
\newcommand{\twoH}         {$\sqrt{s}~=~0.2$~Te\kern-.1emV\xspace}
\newcommand{\twosevensix}  {$\sqrt{s}~=~2.76$~Te\kern-.1emV\xspace}
\newcommand{\five}         {$\sqrt{s}~=~5.02$~Te\kern-.1emV\xspace}
\newcommand{\twosevensixnn}{$\sqrt{s_{\mathrm{NN}}}~=~2.76$~Te\kern-.1emV\xspace}
\newcommand{\fivenn}       {$\sqrt{s_{\mathrm{NN}}}~=~5.02$~Te\kern-.1emV\xspace}

\newcommand{\MeVc}         {Me\kern-.1emV/$c$\xspace}
\newcommand{\TeV}          {Te\kern-.1emV\xspace}
\newcommand{\GeV}          {Ge\kern-.1emV\xspace}
\newcommand{\MeV}          {Me\kern-.1emV\xspace}
\newcommand{\GeVmass}      {Ge\kern-.2emV/$c^2$\xspace}
\newcommand{\MeVmass}      {Me\kern-.2emV/$c^2$\xspace}

\newcommand{\Qsq}{\ensuremath{{Q}^{2}}}

\begin{document}


\title{White Paper on Software Infrastructure for Advanced Nuclear Physics Computing}



\newcommand{\instANL}{Argonne National Laboratory, Lemont, IL 60439, USA}
\newcommand{\instBNL}{Brookhaven National Laboratory, Upton, NY 11973, USA}
\newcommand{\instFRIB}{Facility for Rare Isotope Beams, Michigan State University, East Lansing, MI 48824, USA}
\newcommand{\instJLab}{Thomas Jefferson National Accelerator Facility, Newport News, VA 23606, USA}
\newcommand{\instHPDF}{High Performance Data Facility, Thomas Jefferson National Accelerator Facility, Newport News, VA 23606, USA}
\newcommand{\instLANL}{Los Alamos National Laboratory, Los Alamos, NM 87545, USA}
\newcommand{\instLBNL}{Lawrence Berkeley National Laboratory, Berkeley, CA 94720, USA}
\newcommand{\instLLNL}{Lawrence Livermore National Laboratory, Livermore, CA 94551, USA}
\newcommand{\instORNL}{Oak Ridge National Laboratory, Oak Ridge, TN 37831, USA}


\newcommand{\instFIU}{Florida International University, Miami FL 33199, USA}
\newcommand{\instUCB}{University of California, Berkeley, CA 94270, USA}
\newcommand{\instUH}{University of Houston, Houston, TX 77204, USA}
\newcommand{\instUIUC}{University of Illinois Urbana-Champaign, Urbana, IL 61801, USA}   
\newcommand{\instMSUPhys}{Department of Physics and Astronomy, Michigan State University, East Lansing, MI 48824, USA}
\newcommand{\instMSUChem}{Department of Chemistry, Michigan State University, East Lansing, MI 48824, USA}
\newcommand{\instOldDom}{Old Dominion University, Norfolk VA, 23529, USA}
\newcommand{\instURegina}{University of Regina, Regina, Saskatchewan S4S 0A2, Canada}
\newcommand{\instTemple}{Temple University, Philadelphia PA 19122, USA}
\newcommand{\instUTenn}{University of Tennessee, Knoxville, TN 37830, USA}
\newcommand{\instVTech}{Virginia Tech, Blacksburg VA 24061, USA}
\newcommand{\instUWash}{University of Washington, Seattle, WA  98195, USA}
\newcommand{\instWandM}{College of William \& Mary, Williamsburg, VA 23185, USA}
\newcommand{\instWSU}{Wayne State University, Detroit, MI 48202, USA}


\newcommand{\compRadiaSoft}{RadiaSoft LLC, Boulder CO 80301, USA}


\author{P.~M.~Jacobs\footnote{Corresponding Author: \href{mailto:pmjacobs@lbl.gov}{pmjacobs@lbl.gov}} }
    \affiliation{\instLBNL}
\author{A.~Boehnlein}
    \affiliation{\instJLab} 
\author{B.~Sawatzky}
    \affiliation{\instJLab}
\author{J.~Carlson}
    \affiliation{\instLANL}
\author{I.~Cloet}
    \affiliation{\instANL}
\author{M.~Diefenthaler}
    \affiliation{\instJLab}
\author{R.~G.~Edwards}
    \affiliation{\instJLab}
\author{K.~Godbey}
    \affiliation{\instFRIB}
    \affiliation{\instMSUPhys}
\author{W.~R.~Hix}
    \affiliation{\instORNL}
    \affiliation{\instUTenn}
\author{K.~Orginos}
    \affiliation{\instWandM}
\author{T.~Papenbrock}
    \affiliation{\instORNL}
    \affiliation{\instUTenn}
\author{M.~Ploskon}
    \affiliation{\instLBNL}
\author{C.~Ratti}
    \affiliation{\instUH}
\author{R.~Soltz}
    \affiliation{\instLLNL}
\author{T.~Wenaus}
    \affiliation{\instBNL}
\author{L.~Andreoli}
    \affiliation{\instJLab} 
    \affiliation{\instOldDom}
\author{J.~Brodsky}
    \affiliation{\instLLNL}
\author{D.~Brown}
    \affiliation{\instBNL}
\author{A. Bulgac}
    \affiliation{\instUWash}
\author{G.~D.~Chung}
    \affiliation{\instVTech}   
\author{S.J.~Coleman}
    \affiliation{\compRadiaSoft}
\author{J. Detwiler}
    \affiliation{\instUWash}
\author {A.~Dubey}
    \affiliation{\instANL}
\author{R.~Ehlers}
    \affiliation{\instLBNL}
    \affiliation{\instUCB}
\author{S.~Gandolfi}
    \affiliation{\instLANL}
\author{G.~Heyes}
    \affiliation{\instHPDF}
\author{G.~R.~Jansen}
    \affiliation{\instORNL} 
\author{F.~Jonas}
    \affiliation{\instLBNL}
\author{S.~R.~Klein}
    \affiliation{\instLBNL}
\author{R.~Kr\"{u}cken}
    \affiliation{\instLBNL}   
\author{D.~Lee}
    \affiliation{\instFRIB}
    \affiliation{\instMSUPhys} 
\author{S.~N.~Liddick}
    \affiliation{\instFRIB}
    \affiliation{\instMSUChem}
\author{H.-W.~Lin}
    \affiliation{\instMSUPhys} 
\author{A.~Majumder}
    \affiliation{\instWSU}
\author{T.~A.~Manning}
    \affiliation{\instUIUC} 
\author{O.~E.~B.~Messer}
    \affiliation{\instORNL}
\author{H.~Monge-Camacho}
    \affiliation{\instORNL}   
\author{T.~Munson}
    \affiliation{\instANL}
\author{B.~Nachman}
    \affiliation{\instLBNL}
\author{W.~Nazarewicz}
    \affiliation{\instFRIB}
    \affiliation{\instMSUPhys} 
\author{E.G.~Ng}
    \affiliation{\instLBNL}
\author{A.~Panta}
    \affiliation{\instJLab}
\author{J.~Putschke}
    \affiliation{\instWSU}
\author{T.~Reed}
 \affiliation{\instFIU}  
\author{F.~Salazar}
 \affiliation{\instTemple}
\author{N.~Sato}
    \affiliation{\instJLab}
\author{M.~Savage}
    \affiliation{\instUWash}
\author{B.~Schenke}
    \affiliation{\instBNL}
\author{L.~Schwiebert}
    \affiliation{\instWSU}
\author{C.~Shen}
    \affiliation{\instWSU}
\author{G.~Vujanovic}
    \affiliation{\instURegina}
\author{A.~Walker-Loud}
    \affiliation{\instLBNL}

\date{\today}

\begin{abstract}

This White Paper documents the discussion and consensus conclusions of the workshop ``Software Infrastructure for Advanced Nuclear Physics Computing'' (SANPC 24), which was held at Jefferson Lab on June 20--22, 2024. The workshop brought together members of the US Nuclear Physics community with data scientists and funding agency representatives, to discuss the challenges and opportunities in advanced computing for Nuclear Physics in the coming decade. Opportunities for sustainable support and growth are identified, within the context of existing and currently planned DOE and NSF programs.

\end{abstract}



\maketitle


\newpage
\tableofcontents
\setcounter{tocdepth}{3}


\newpage
\section{Executive Summary}
\label{Sect:ExecSum}

Software, computing, and ever-larger datasets have become key drivers of Nuclear Physics (NP) research, due to the rapid growth in scientific facilities, algorithms, and computational capacity. The 2023 Long Range Plan (LRP) for Nuclear Science emphasized the importance of capitalizing on these extraordinary opportunities for scientific discovery, identifying the investment in emerging technologies, advanced computing, and multidisciplinary centers as a high priority for the field of Nuclear Physics.

Federal agency programs, including the DOE Scientific Discovery through Advanced Computing (SciDAC) program, the NSF Cyberinfrastructure for Sustained Scientific Innovation (CSSI) program, and the DOE Exascale Computing Project (ECP), have successfully supported multi-disciplinary collaborations of nuclear scientists, computer scientists, and applied mathematicians to meet this challenge, enabling scientific advances and software innovation. Collaboration with industry on cutting-edge technologies has also proven to be highly effective. Such collaborations will continue to be essential in the future for tackling forefront nuclear physics problems with innovative approaches.

As the next step in realizing the vision for advanced computing expressed in the LRP, a workshop entitled \textit{Software Infrastructure for Advanced Nuclear Physics Computing} (SANPC 24) was held at Jefferson Lab on June 20--22, 2024. This White Paper presents the discussion and consensus of the workshop participants about critical research directions, challenges, and opportunities for sustainable support and growth, to realize the LRP vision. Workshop attendees represented all sub-areas of the US Nuclear Physics portfolio. The Workshop organization and agenda are found in the Appendix.

The primary observations and opportunities identified at the workshop for support of the software infrastructure needed for groundbreaking advances in Nuclear Physics are as follows:

\noindent{\bf Innovation} Sustainable software infrastructure for nuclear physics discovery requires innovation in hardware and algorithmic development, and the inclusion of approaches and best practices from industry. Such innovations are often implemented in common software ecosystems that have broad application in the NP research portfolio. The importance of these ecosystems will increase as they incorporate further innovations in AI/ML and other rapidly developing technologies. \\


\noindent{\bf Cross-cutting initiatives} The NP research enterprise has benefited greatly from multi-disciplinary efforts, which drive innovation by incorporating developments in computer science, applied mathematics, statistics, and from emerging industries. Ongoing partnerships with industry to exploit cutting-edge technologies have proven to be highly effective and cost-efficient. Sustainable support mechanisms are needed for such multi-disciplinary collaborations, with flexibility to support projects of widely varying scale.  \\

\noindent{\bf Stewardship} It is vital that funding instruments enable continuity and evolution of community software and software infrastructure. Sustained support in these areas leverages current investments and promotes software functionality on new HPC and other novel hardware systems. Effective stewardship should cover the full lifecycle of data and software management.  \\ 

\noindent{\bf Data curation and preservation} Robust data and software preservation is essential to maximize the return on current investments in NP experimental facilities and instruments. The development of effective mechanisms for long-term preservation of data and associated metadata, including research software and workflows, is needed to ensure long-term accessibility and reproducibility of publicly--funded science. \\

\noindent{\bf Talent retention} Forefront NP research depends upon high--performance computing and large-scale data analysis. Retention of an expert workforce in this area requires sustained funding support and viable academic career paths for scientific software developers working at the intersection of nuclear physics, data science, and computing. Retention mechanisms include mentorship programs, training opportunities, and the recognition of software contributions as integral to scientific research.

\newpage
\section{Introduction}
\label{Sect:Intro}

Software and computing have become drivers in Nuclear Physics (NP), due to the rapid growth of scientific data and exascale computing facilities. Exploitation of these opportunities requires new hardware architectures and new algorithmic approaches at exascale, including those based on Artificial Intelligence and Machine Learning. 

New collaborative efforts have emerged to capitalize on these opportunities, bringing NP theorists and experimentalists together with 
data scientists, computer scientists, and applied mathematicians. A common theme of such collaborative efforts is the development of robust, large-scale software infrastructure frameworks to carry out their complex scientific programs. While these frameworks require long-term support, the establishment of suitable support mechanisms for that purpose is challenging. The curation and long-term preservation of data and associated analysis tools is an important related issue. 

The 2023 Long Range Plan of the Nuclear Science Advisory Committee (NSAC LRP)~\cite{NSACLRP} incorporates a vision of the future of advanced computing for NP research. A workshop entitled ``Software Infrastructure for Advanced Nuclear Physics Computing'' (SANPC)~\cite{SANPC} was held at Jefferson Lab on June 20--22, 2024, bringing together NP computing practitioners to discuss the implementation of this vision, with a focus on identifying mechanisms to achieve long-term, sustainable software infrastructure. The workshop program covered a broad range of current software and computing projects in the NP domain. Discussions addressed the role of research vs. infrastructure software, development of common software ecosystems, software sustainability best practices, and workforce development. Existing and currently planned DOE and NSF programs which support advanced NP computing and cyber-infrastructure were reviewed.

Nuclear Physics is by its nature data-based, both in experiment and in many areas of theory and modeling. The development of software for advanced computing is therefore driven to a large extent by the requirements of data management and analysis, and the interplay between data and software was a common theme in workshop discussions.

This White Paper presents the workshop discussion and consensus about critical research directions, challenges, and opportunities for Advanced Computing in Nuclear Physics. Similar considerations can be found in the White Paper from an earlier workshop on the application of AI to NP research~\cite{Bedaque:2021bja}.

A list of acronyms used in the report can be found in Sect.~\ref{sect:Acronyms}. Additional material can be found in the slides of the workshop presentations~\cite{SANPCindico}.

\section{Observations and opportunities}
\label{Sect:ObsOpp}

This section summarizes the observations made at the workshop about the key issues affecting advanced computing efforts in NP, and the opportunities identified to address them. The supporting material for this discussion is presented in Sect.~\ref{Sect:CurrentLandscape}. Capitalizing on these opportunities can transform NP research through the adoption of new technologies and the expansion of an expert, highly skilled workforce.

The NP advanced computing landscape sketched in Sect.~\ref{Sect:CurrentLandscape} is wide-ranging, encompassing projects that require large teams to address broad topics in theory, experiment, and their intersection, and small teams focusing on more limited, targeted questions. NP projects are carried out at leading accelerator facilities, at high-performance computing facilities, and on small, purpose-built experimental equipment and computing systems. The needs and requirements for the support and development of advanced NP computing are consequently highly diverse, with no single set of proposed solutions applicable to the entire community.

In order to address this diversity of needs and requirements, discussion at the workshop was carried out in part in three parallel Working Groups, focusing on theory, on experiment, and on joint theory/experimental collaborations. Each Working Group reported the results of their deliberations in a follow-up plenary session, and in written documentation. Despite the different scientific scope of the three Working Groups, many observations and proposed solutions were found to be common among them. This section therefore presents these observations and proposed solutions in a unified way. 

Many aspects of the workshop discussion and conclusions address the FAIR principles which have been proposed as guidance for all scientific data management and stewardship~\cite{Wilkinson:2016myn}: Findability, Accessibility, Interoperability, and Reusability. The overlap of workshop discussion with FAIR principles is widespread in the following, and we do not call out each instance.

\subsection{Community organization} 

\subsubsection{Opportunities}

A strong consensus at the workshop supported the establishment of a standing committee representing the broad NP computing community, to coordinate discussion and provide guidance on matters related to advanced NP computing. This committee could also be charged with organizing periodic meetings on advanced NP computing, perhaps similar in nature to this workshop. The NP computational community comprises a wide range of organized collaborations together with smaller, less formal efforts, with differing needs for support. The scope of discussions at such meetings should include software stewardship, workforce development, cross-project support and coordination, and data curation and preservation. 

\subsection{Software stewardship}

\subsubsection{Observations}

``Stewardship'' refers to the planning and management of resources. Within the NP community there is a large body of community software and infrastructure, representing significant investment over many years, that are valuable assets which need to be preserved. 

Stewardship is needed to maintain key libraries in widespread community use. Such maintenance includes the integration of new features and elements into existing software, and the adaption of software to run on new HPC platforms. As an example, following completion of the Exascale Computing Project~\cite{Exascale}, DOE ASCR is providing support for software stewardship of codes and libraries that it considers to be essential for its ongoing scientific efforts and computing facilities. 

Stewardship provides users with the assurance of continued availability, thereby reducing the duplication of effort and of code bases, and standardizing software citations as recognition and incentive for code authors. 

Stewardship also enables the evolution of applications, providing a development path for functionality on emerging HPC and other systems.  Stewardship of existing software therefore complements the development of new capabilities.

Numerous applications of AI/ML to NP are currently being developed and will need stewardship support. In the case of Quantum Information Science, industry employs significantly more resources to support such efforts than universities and national laboratories. However, such support can be volatile, and funding agencies have a role in supporting the preservation of such efforts beyond the lifetime of industry investments.

Sustained stewardship also provides more stable career pathways for the domain-science oriented software engineers, who are essential to NP research areas which rely on HPC resources and advanced technologies.

\subsubsection{Opportunities}

Sustained support for software stewardship is vital for the NP community to fully utilize current and future HPC and other computing resources, and to maximize their scientific output at manageable cost. 

Resources for software stewardship are required to ensure that existing software can adapt to new platforms and technological advances, notably AI/ML and quantum computing.  Similarly, sustained support is needed for open-source software, especially frameworks developed by large collaborations, beyond their nominal period of operation.

Institutional computing centers play an important role in the development and stewardship of common software ecosystems, and should ensure that the barrier to entry is low and that software information and support are readily available.

\subsection{Computing scope and resources}

\subsubsection{Observations}

Nuclear Physics experimental efforts vary widely in scale, from large collaborations at major experimental facilities to small university-based programs. NP theory efforts likewise vary widely in scale, from sizable collaborations focusing on lattice Quantum Chromo--dynamics (QCD), modeling of the Quark-Gluon Plasma, or neutron star dynamics, to few-author calculations. Joint collaborations of theorists, experimentalists, and computing and data scientists are typically on the larger side. 

The size and scope of a collaboration are correlated with its capabilities for software development and maintenance. Large labs have access to significant software engineering effort and resources, to build robust software infrastructure. In contrast, smaller collaborations with fewer resources may not be able to develop and maintain adequate software solutions themselves, due to limited access to expertise and funding. Ensuring such support is crucial to ensure that small experiments can achieve their scientific objectives.

Reproducibility and replicability are crucial aspects of the scientific process, to ensure that research findings are reliable and can be independently verified. Nuclear physics research, both experimental and theoretical, often involves complex, multi-step workflows, with multiple stages of data manipulation and analysis. Such complexity can make it challenging to track the origin and evolution of data and code, obscuring the path taken from raw data to final results. Open-source analysis code and workflows can help in this regard. However, analysis procedures often incorporate undocumented assumptions or rely on internal knowledge - information that is passed down within a research group but not formally recorded - which can result in misinterpretation or difficulty in replicating or reinterpreting the work. The absence of expert knowledge is a significant barrier to analysis transparency and reproducibility. 

\subsubsection{Opportunities}

Good practices to address these challenges include version control systems for code, cataloged documentation of analysis procedures, and the development of standardized data formats and analysis pipelines. Containerization technologies such as Docker can also help to encapsulate software dependencies and ensure reproducibility across different computing environments.

For software to be utilized effectively, it must be easily discoverable. Creating a comprehensive catalog of available software used within the NP community would facilitate access and enhance user engagement.
    
Strengthening partnerships between academic institutions and larger laboratories can help small experimental programs gain access to vital software engineering talent and resources. 

The data generated by various projects also differ in size and complexity, with large projects producing vast amounts of data that require sophisticated data management and analysis pipelines, and therefore a flexible and scalable approach to software development. There is ample opportunity in this area for knowledge-sharing and collaboration among experiments, both large and small. The establishment of common practices and frameworks can enhance software development across the board, and foster a culture of collaboration.

Advanced computing projects commonly require long-term HPC resource allocation, using mechanisms such as INCITE, ERCAP/NERSC, Jetstream2/Indiana, Access/NSF, and ALCC/ASCR. Optimally, funding for project effort should be accompanied by an HPC allocation, which then does not need to be applied for separately. 

\subsection{Innovation and crosscutting initiatives}

\subsubsection{Observations}

The establishment of sustainable software infrastructure for NP discovery requires innovation in hardware and algorithmic developments, and the adoption of approaches and best practices from industry. These innovations are often implemented in software ecosystems with broad application in the NP research portfolio, whose utilization is expected to increase in future. 

Prominent examples include developments in advanced nuclear theory algorithms and in AI/ML and Quantum Computing. Accelerated computing at mixed and reduced precision is a promising avenue for software solutions in various areas, expanding the range of scientific questions that can be addressed while strengthening the connection between NP and the AI/ML hardware industry. However, further R\&D is required for its practical implementation.

The NP research enterprise has benefited significantly from multi-disciplinary efforts, such as those supported by SciDAC, which combine innovations in computer science, applied mathematics, statistics, and emerging industries. Such projects attract experts in a broad range of scientific and technical domains, both within and beyond traditional NP boundaries, fostering partnerships between academic fields and between academia and industry.

As an example, the growing demand for NP computing requires the development novel computer languages for optimal utilization of silicon-based computing resources, leveraging both traditional scalar and vector computing in CPUs and GPUs, and accelerated tensor-based computing in use within AI/ML silicon. This multi-disciplinary effort brings together ideas in physics, mathematics, statistics and computer science.

Partnerships of NP advanced computing projects with industry are crucial for these developments. For classical computing, collaboration with companies such as NVIDIA, Intel, and IBM provides access to state-of-the-art GPU-accelerated machines. Collaboration with industry will benefit language and compiler/interpreter development for silicon-based computing. For NP quantum computing, ongoing projects include collaboration with IBM, Microsoft, IonQ, and Quera.

The Exascale Computing Project (ECP)~\cite{Exascale} demonstrates the benefits of software and architectural {\em Co-Design}. The ECP project, which formally ended in Dec. 2023, funded {\em Path-Forward} efforts to establish new HPC hardware capabilities. ECP participants, including computer scientists, applied mathematicians and domain scientists, collaborated with industry by providing computational and software requirements for scientific campaigns. The industrial partners optimized their tool chains and the performance of community--developed software packages to meet these requirements. Lessons learned from this collaboration have been used to improve industrial products and the performance of the software packages. As new computational platforms come online, it is vital for the community to maintain such industrial partnerships into order to fully exploit the scientific potential of HPC facilities for NP discovery.

Another productive collaboration between NP researchers and industry is the development of a GPU-powered online reconstruction system for the ALICE experiment at the CERN Large Hadron Collider~\cite{ALICE}. For this project, ALICE scientists worked in partnership with Advanced Micro Devices (AMD) on  hardware design and co-development of sophisticated synchronous data reconstruction software and workflows. This deep technical collaboration, which included joint work on compiler and GPU-targeted optimization, was mutually beneficial, resulting in an efficient, production-ready hardware system and a stress-tested software toolkit for AMD GPU systems. This project is a prototype for future NP/industry collaboration in hardware design and software development, especially for data-intensive NP projects such as ALICE.

The Small Business Innovation Research (SBIR) program plays an important role in fostering collaboration between NP researchers and industry, funding modest-scale R{\&}D in key forefront technologies and practices.

\subsubsection{Opportunities}

Multi-disciplinary programs such as SciDAC are highly beneficial to the NP research enterprise. Additional flexibility in such programs, in terms of scope and size of award, would expand their reach and impact.

Close collaboration with industry has proven to be highly productive, cost-effective, and mutually beneficial. Such collaborations should be actively encouraged by mechanisms supporting sustained partnerships.

The workshop consensus supports the recommendations of the 2023 NSAC Long Range Plan for a multi-institution Quantum Computing Center, and a coordinated effort to incorporate AI/ML techniques into NP.

\subsection{Software development and common tools}

\subsubsection{Observations}

NP research encompasses theoretical, experimental, and joint theory/experiment efforts with a broad range in scope and size. Some collaborations are funded specifically to develop software-as-a-service to the community, while others are primarily science-driven, with the software a by-product to some degree. A collaborative approach is beneficial in both cases, enabling sustained discussion, organized distribution of tasks, and sharing of knowledge. Computing-oriented nuclear physicists and research software engineers are crucial elements of such multi-disciplinary projects. 

The application of AI/ML can significantly improve software efficiency, reproducibility, and interpretability. Physics-driven AI/ML algorithms enhance both data analysis and decision-making processes. 

During the workshop, several common techniques were identified among NP projects, including Bayesian inference and other inverse-problem solutions, Machine Learning applications, adaptive grids, and numerical linear algebra. Common issues include uncertainty quantification, data storage and sharing, and software development and stewardship. Common tools include Kokkos, Docker, and QUDA. 

\subsubsection{Opportunities}

Software development and operations in large projects should not be regarded as separate efforts. Collaboration between these teams provides continuous improvement and immediate feedback, promoting software robustness and functionality.

The integration of innovative tools developed outside the traditional NP community can enhance software capabilities. Multi-disciplinary collaborations can achieve innovative solutions by reusing existing tools and adapting them to NP applications. 

Common software tools should be modular and portable, using common standards and with meaningful documentation. 

\subsection{Data and analysis preservation}

\subsubsection{Observations}

Curation and preservation of data and analysis are essential to ensure the longevity and reproducibility of scientific research. As the complexity of scientific workflows grows, the re-analysis of data and reproducibility of results become significant challenges. For many projects, the focus during their active phase is primarily on immediate physics results, rather than on long-term archival needs. After completion of the active project phase, analysis and simulation tools tend to become black boxes, with insufficient documentation and complicated dependencies, making reliable data re-analysis difficult or impossible. To address this challenge it is necessary to preserve raw data and their associated software, workflows, calibration metadata, and ML models~\cite{LHCReinterpretationForum:2025zgq}. The infrastructure to support this task must allow both for long-term storage and for future usability of these resources. 

There is an additional, less tangible but no less important, element to data preservation, which is the preservation of institutional knowledge. There have been recent efforts to re-analyze archived data from the LEP $e^+e^-$ collider at CERN~\cite{Badea:2019vey,Chen:2021uws,Chen:2023njr} , which terminated operation in 2000, and the HERA ep Collider at DESY~\cite{Bacchetta:2016rdn,H1:2024pvu,H1:2024aze,H1:2024nde,H1:2023fzk,H1:2021wkz,Malka:2012hg,Verbytskyi:2016vtw,Diefenthaler:2021rdj}, which terminated operation in 2008. In addition, the large LHC collaborations have released open data from early LHC runs, with CMS collaboration data analyzed by external groups~\cite{Larkoski:2017bvj,Komiske:2019jim}. These collider detectors, and their datasets and analysis procedures, are too complex for the full documentation of all essential details in a comprehensible way. Rather, practical experience with such re-analyses has revealed that bringing them to publication-level quality requires expert knowledge from physicists who have participated in experimental operations and analysis during the active phase of the project. This is an important in-practice lesson for data curation and preservation projects: expert knowledge also needs to be preserved. The absence of well-defined career paths for scientists focused on the development and maintenance of domain-specific software therefore presents a challenge for data preservation. 

Mechanisms for the curation and preservation of data were found to be common to experimental and theoretical efforts in NP. As experimental and computational capabilities have developed, the consideration of policies governing data stewardship has become more complex. Stewardship and preservation of AI/ML methods in analysis workflows is especially challenging, given their dependence on training data and model construction.

Workshop discussion identified specific challenges for data and analysis preservation:

\begin{itemize}
    \item Infrastructure needs: Long-term infrastructure is necessary to store data, software, and metadata, ensuring that archived workflows can still be executed in the future.
    \item Standardization: The lack of standardized and well-documented formats for storing data and workflows presents a barrier to reproducibility and collaboration.
    \item Discoverability: Even if data and workflows are preserved, ensuring they are easily discoverable by future researchers needs attention.
    \item Centralized support: Tasks such as the maintenance of archiving infrastructure and the implementation of best-practice policies require centralized efforts and community-wide coordination.
    \item Software lifecycle planning: Sustained support is vital for software lifecycle management and planning, including updates and implementation on new platforms.
    \item Proprietary Software/Hardware: R\&D efforts outside of NP, including industry, play a critical role in cutting-edge endeavors such as AI/ML and Quantum computing. Such external organizations provide valuable resources to support the NP research. It is important to maintain such resources (proprietary tool chains, compilers, libraries, cloud services), whose disappearance could render large datasets unrecoverable or uninterpretable.
\end{itemize}

\subsubsection{Opportunities}

Support is vital for the development and maintenance of robust frameworks for the preservation of data and associated metadata, including research software and workflows, to ensure their long-term accessibility and the reproducibility of scientific results.
 
Investment in decentralized storage networks would be valuable, thereby leveraging and extending mature technologies such as IPFS~\cite{IPFS} to flexibly distribute responsibility and to lower the cost of long-term storage of scientific data.    

\subsection{New facilities}

\subsubsection{Observations}

The US Department of Energy has initiated the Integrated Research Infrastructure (IRI) Program to integrate its experimental facilities, data assets, and advanced computing resources. 

The High Performance Data Facility (HPDF) Project\cite{HPDF} has been established to address gaps in data curation, preservation, and time-critical discovery at existing ASCR facilities. HPDF will become the fifth ASCR facility, designed as a distributed platform with a resilient hub across two geographic locations, offering stable data services like lifecycle management, visualization, search, and publishing. Unlike traditional ASCR facilities focused on computing, HPDF will emphasize the value of scientific data through co-designed solutions with user communities and training in data stewardship.

This data-centric approach aligns with four of the five key opportunities identified in this White Paper: fostering innovation (developing software ecosystems, supporting AI/ML, and emerging technologies), enabling cross-disciplinary initiatives, advancing data curation and preservation, and promoting software stewardship. Close collaboration of the NP community and HPDF is essential to achieve these goals.

\subsection{Career paths and talent retention}

\subsubsection{Observations}

Nuclear Physics is increasingly dependent upon high-performance computing and large-scale data analysis to address forefront research problems. Data physicists provide specialized skills in data management and curation, algorithm development, and software engineering. Scientific software developers drive innovation through the adoption of new technologies such as AI/ML, and support operations. 

However, for early-career nuclear physicists with interest in academic positions, it is challenging at present to focus on software development, since hiring metrics for faculty positions are based primarily on publications, citations, and presentations at physics conferences, while software development is not recognized at the same level. The support of viable career paths and incentive structures for scientific software developers at the intersection of nuclear physics, data science, and computing is critical for retaining and developing talent within the field. Retention mechanisms for such staff could include mentorship programs, training opportunities, and recognition of software contributions as integral to scientific research.   

It is likewise important to train the next generation of computing-focused scientists, fostering students who show an interest and aptitude in this area. A strong foundation in data science and computational science includes expertise in data manipulation, analysis, and visualization techniques, as well as machine learning and uncertainty quantification. These skills are also valued in the broader workforce, enabling successful careers outside of academia.
 
\subsubsection{Opportunities}

The workshop consensus advocated for the support of long-term positions for nuclear physicists, both theorists and experimentalists, with the skills and interest in the development and stewardship of software.

The field should acknowledge prominently that software and computing are essential components of many projects and experiments, to help ensure that they receive adequate funding. This includes the emphasis in project proposals on analysis methodology and computing resources. 

An organization focused on software development and stewardship in NP could develop and disseminate training modules and video tutorials, and run online training and discussion sessions, and hackathons. 

It is also important to recognize that expertise outside NP is valuable for NP research. Fostering such connections requires the promotion of multi-disciplinary opportunities and the recruitment of expertise beyond conventional NP boundaries. Partnerships with industry are beneficial to the NP research program and to the training of students and postdocs, opening up future employment opportunities for them.

\section{Supporting material}
\label{Sect:CurrentLandscape}

This section provides context for the workshop discussion in terms of the current funding instruments and coordinating organizations, an overview of advanced NP computing in theory and experiment, and a survey of current grant-funded projects. Additional workshop presentations on selected topics are also described.

\subsection{Funding instruments}
\label{Sect:FundingInstruments}

Sustained funding for the core effort in advanced computing in NP is provided by the Nuclear Physics program in the DOE Office of Science~\cite{DOENP} and the Division of Physics in the Mathematical and Physical Sciences directorate of the National Science Foundation~\cite{NSFNP}. Additional funding by time-limited grants is provided by various agencies, as follows.

\subsubsection{ASCR}
\label{secASCR}

The mission of the Advanced Scientific Computing Research program (ASCR)~\cite{ASCR} is to develop, deploy, and utilize US high-performance computational and networking capabilities to analyze, model, simulate, and predict complex phenomena. Special emphasis is placed on emerging computing systems and novel computing architectures. 


ASCR has recently launched a new partnership with the NNSA and the Linux Foundation under the High Performance Software Foundation (HPSF)~\cite{HPSF}, which is aligned with the ASCR mission to support High-Performance Computing (HPC) systems. HPSF aims to build, promote, and advance a portable software stack for HPC systems by increasing adoption, lowering barriers to contribution, and supporting development efforts.

Additional seed projects~\cite{ASCRseedprojects} have been constituted to develop a sustainable software ecosystem for Post-Exascale Computing (PEC), organized under the umbrella of the Consortium for the Advancement of Scientific Software~\cite{CASS}:  Collaboration for Better Software~\cite{COLABS}; Center for Open-Source Research Software Advancement~\cite{CORSA}; Partnering for Scientific-Software Ecosystem Opportunities~\cite{PESO}; Stewardship for Programming Systems and Tools~\cite{S4PST}; Software Tools Ecosystem Project~\cite{STEP}; and Stewardship and Advancement of Workflows and Application Services~\cite{SWAS}.

\subsubsection{NP-SciDAC partnership}
\label{SciDACintro}

The SciDAC program (Scientific Discovery Through Advanced Computing, \cite{SciDAC}) develops the scientific computing software and hardware infrastructure needed to advance scientific discovery using supercomputers. SciDAC is a partnership of several DOE Office of Science programs, including Nuclear Physics. The SciDAC program incorporates two institutes: FASTMath~\cite{FASTMath}, which develops algorithms, AI/ML capabilities, and software tools for simulation and analysis of complex physical phenomena; and  RAPIDS2~\cite{RAPIDS}, which addresses computer science, data, and AI challenges in the utilization of supercomputing resources. Applied mathematicians and computer scientists from these institutes collaborate with the domain scientists on projects of common interest.

SciDAC provides significant funding for NP projects focusing on exascale computing, through multi-disciplinary partnerships. 


\subsubsection{CSSI/NSF}
\label{secCSSI}

The Cyberinfrastructure for Sustained Scientific Innovation program (CSSI)~\cite{CSSI} of the National Science Foundation supports multi-disciplinary (multi-directorate) projects in the development of cyberinfrastructure (CI), emphasizing integrated CI services, quantitative metrics of delivery and usage, and community creation. There are three classes of awards:

\begin{itemize}

\item Elements: small-group awards for the creation and deployment of services for a targeted need. 
    
\item Framework Implementation: larger, interdisciplinary team awards to develop and apply services supporting a broad research area in the NSF science and engineering portfolio, providing a sustainable software framework to a diverse community.
    
\item Transition to Sustainability: awards for groups to provide sustained support of existing CI that will impact a broad research area in the NSF science and engineering portfolio, using new methodologies.

\end{itemize}
    
Support for Elements and Frameworks is of finite duration, up to 10 years.  NSF funding is generally only available to university groups.

\subsubsection{DNN/NNSA}
\label{secDNN}

The Office of Defense Nuclear Nonproliferation (DNN)~\cite{DNN}, in the DOE National Nuclear Security Administration (NNSA)~\cite{NNSA}, supports R\&D in data analysis and AI for nuclear non-proliferation and nuclear security. Current research areas include the modeling of complex systems, measurements with sparse and noisy data, and reliable detection of rare events. These challenges drive investment in AI technologies such as foundational models, knowledge graphs, and AI agents, and benchmarking for testing and evaluation.

The DNN supports several applied NP projects.

\subsubsection{Simons Foundation}
\label{Simons}

The Simons Foundation Scientific Software Research Faculty Award~\cite{Simons} supports scientific software-focused research professor positions in existing academic programs, by providing 50\% support. The goal of this program is to stimulate the development and maintenance of core scientific software infrastructure in academic environments by creating a new, long-term, faculty-level career path. There are a limited number of such awards each year.


\subsection{Advanced computing in NP: theory}
\label{Sect:ACtheory}

Advanced computing plays a central role in NP Theory. This section sketches the advanced computing landscape in the major NP theory research areas.

\subsubsection{Lattice QCD}

Quantum Chromodynamics (QCD) is by its nature non-linear because the gluon, the QCD force moderator, carries color charge and interacts with other gluons. The non-linear nature of QCD results in rich phenomenology, including the confinement of quarks and  gluons in hadrons. However, theoretical calculations of QCD phenomena are complex, requiring specialized approaches. 

Lattice QCD (LQCD), a formulation of QCD in Euclidean spacetime, is amenable to numerical methods and it remains the only way to compute the properties and interactions of nuclear matter directly from QCD with fully quantifiable and improvable systematic uncertainties~\cite{LatticeQCD}.
Through SciDAC and other partnerships, the US LQCD community has developed a number of software libraries that have enabled extensive calculations by the broad community on both capacity and leadership--class computing.
LQCD calculations are particularly well suited to GPUs.
As such, LQCD underwent a disruptive acceleration with the advent of GPUs, of which our community was an early user~\cite{Barros:2008rd,Clark:2009wm}.  Since then a large fraction of the LQCD workflow has been ported to GPUs, which dominate available exascale computing resources.

LQCD, which is formulated in four dimensions with complex--valued fields, is computing-intensive. Its features are somewhat unique in the HPC world, and a large fraction of the core LQCD software libraries have been developed by the LQCD community itself. The LQCD community has consequently become proficient in state-of-the-art algorithms and the efficient utilization of hardware and communication infrastructure.
Several members of the LQCD community have moved to industry, leading to strong partnerships in software development with AMD, Intel and NVIDIA. 

SciDAC and similar initiatives enable the LQCD community to maintain and build partnerships with industry and with colleagues in computer science and applied mathematics, which are vital for the future success of LQCD research. In particular, these partnerships focus on making LQCD software more \textit{performance--portable}. They also help to develop and implement state-of-the-art algorithms for optimizing performance, such as the multi-grid algorithm for solving for the inverse of large sparse matrices~\cite{Clark:2016rdz}.

Machine Learning--based methods also offer new opportunities for developing efficient algorithms for LQCD, which benefit from collaboration with ML experts. New leadership computing facilities are poised to incorporate AI/ML-specific nodes in the computing infrastructure, and it is important to ensure that LQCD libraries can make use of these resources, to maximize the scientific output.

\subsubsection{Partonic structure of hadrons}
\label{Sect:partonstructure}

The partonic structure of hadrons is of fundamental importance in physics, and its experimental and theoretical study are at the forefront of modern research in Quantum Chromodynamics (QCD). Theoretical QCD calculations predict with precision how partonic structure changes with scale (``QCD evolution''), connecting phenomena that span several orders of magnitude and bridging experimental observations from different colliders. The evolution with respect to momentum transfer \Qsq, which is dictated by the Dokshitzer-Gribov-Lipatov-Altarelli-Parisi (DGLAP) equation~\cite{Gribov:1972ri,Lipatov:1974qm,Altarelli:1977zs,Dokshitzer:1977sg}, has been tested extensively, providing compelling evidence for QCD as the correct theory of strong interactions.

Equally important but less understood is the evolution of hadronic structure with respect to rapidity (or equivalently Bjorken $x$, the longitudinal momentum fraction). The Balitsky-Fadin-Kuraev-Lipatov (BFKL) equation~\cite{Lipatov:1976zz,Kuraev:1977fs,Balitsky:1978ic}, which governs small-$x$ evolution at high center-of-mass energy for fixed \Qsq, predicts a steep rise in parton densities toward small $x$~\cite{Ball:2017otu}. However, when parton densities reach a critical threshold, a novel phenomenon known as gluon saturation is expected to emerge, where gluon recombination counterbalances further growth \cite{Gelis:2010nm,Kovchegov:2012mbw,Albacete:2014fwa,Morreale:2021pnn}. In this regime, rapidity evolution is no longer governed by linear BFKL dynamics but by nonlinear equations which account for gluon saturation effects at high parton densities, such as the Balitsky-Kovchegov (BK) equation \cite{Balitsky:1995ub,Kovchegov:1999yj}, or more generally, the Jalilian-Marian–Iancu–McLerran–Weigert–Leonidov–Kovner (JIMWLK) equations 
\cite{Jalilian-Marian:1996mkd,Jalilian-Marian:1997ubg,Kovner:2000pt,Iancu:2000hn,Iancu:2001ad,Ferreiro:2001qy}. 

The BK equation is a non-linear integro-differential equation, making its numerical computation complex and expensive. This poses a significant challenge for global phenomenological analyses comparing QCD calculations with a broad array of data in the search for non-linear QCD evolution effects. Novel machine learning tools such as Gaussian emulators and generative AI models are being developed to reduce the complexity of calculations with the BK equation; this is particularly important for solving the next-to-leading order (NLO) BK equation \cite{Lappi:2015fma}, which is under development for the high-precision analysis of future data from the LHC and EIC. The more complete JIMWLK evolution is a hierarchy of coupled non-linear integro-differential equations, making its numerical computation yet more complex and expensive. At leading logarithmic order it has been brought into a Langevin form \cite{Rummukainen:2003ns} that is computationally tractable at existing facilities. This is typically carried out using a two-dimensional spatial lattice which is numerically intensive, but efficient algorithms such as fast Fourier transforms render the numerical effort comparable to that needed to solve the BK equation. Formulation of the NLO logarithmic version of the JIMWLK equation, in a way that is amenable to numerical solution, is under development.

\subsubsection{Nuclear properties}

Accurate calculations of nuclear properties, including structure, nuclear fission and fusion, electro-weak transitions, and response to electron- and neutrino-scattering, require extensive computing resources. Significant methodological advances for the nuclear quantum many-body problem have been made over the past two decades, including Quantum Monte Carlo methods (continuum or lattice), Coupled Cluster theory, continuum shell models, and density functional theory.

Such calculations can require the largest available supercomputers, with large memory and GPU usage. Current codes for such calculations have been developed primarily by nuclear scientists, whose background in usually not in computing science, and their adaptation to complex new computing architecture and libraries has proven to be challenging. It is crucial to modernize these codes, in order to carry out calculations in new physical regimes in mass and dynamical time scales. The improvement of legacy codes, the development new codes, and the deployment of advanced  AI/ML approaches for the calculation of nuclear properties requires the extensive collaboration of nuclear physicists, computer scientists, and data scientists. As outlined in Sect.~\ref{Sect:FundedProjects}, several such collaborations are ongoing, employing for instance efficient ML-based emulators and neural networks to improve the many-body wave function.

The future development of nuclear many-body methods will require continued and expanded support of such cross-cutting collaborations for the effective usage of new hardware resources architectures, and continuously evolving AI/ML libraries and methods.

\subsubsection{Nuclear astrophysics}

The lives and deaths of stars play an essential role in the origin of the chemical elements~\cite{Arcones:2023,Johnson:2020}.
We are isolated from the stellar events that are responsible for this nucleosynthesis by distances measured in kiloparsecs and times measured in billions of years.  
Simulations reveal the inner workings of these nucleosynthesis events and disclose details that are otherwise hidden by distance and time, enabling us to deduce what happens as stars explode and nuclei are made.  
Among the details that such simulations reveal are key nuclear reaction pathways. This knowledge helps guide programmatic planning to maximize the scientific gain from experiments at FRIB and other nuclear experimental facilities~\cite{Schatz:2022}. 

To study each site of astrophysical nucleosynthesis, a chain of simulation, with links implementing different physics, is used to build the event progenitor, witness the event, calculate the nucleosynthesis that the event produces, and follow the newly-formed nuclei until they reveal themselves to telescopic observation. \cite{Burns:2025} Predictions for the event and its nucleosynthesis can be tested by comparing to observations that follow days or years after the event.  
Such simulations are computationally expensive, requiring significant resources at HPC facilities for accurate predictions \cite{Carlson:2017}.
Achieving sufficiently high physical fidelity requires high resolution, in both coordinate and phase space.  
At the same time, the wide variety of physical processes at work simultaneously (radiation transport, strong, electromagnetic and weak nuclear reactions, etc.), each with its accompanying timescale, generally renders the system of equations stiff and therefore expensive to integrate over time.
Continuous improvement enables simulations with greater physical fidelity, higher resolution and longer duration over the past decade.
Significant developments have taken advantage of exascale resources, which required the extensive collaboration between physicists, computer scientists and applied mathematicians.
Further computational improvements are essential to understand the wealth of observations coming in the next decade from time-domain and multi-messenger astronomy \cite{Burns:2025}.   

Within the next decade we expect that the combination of high-performance computing and AI/ML will enable dramatic improvements in our understanding of the nuclear equation of state and neutrino flavor oscillations.
These areas are themselves key elements of the US NP research portfolio, and their deeper understanding will provide essential input to simulations of core-collapse supernovae and neutron star mergers, and to our understanding of nucleosynthesis.


\subsubsection{Heavy-ion physics}

Experimental measurements of relativistic heavy-ion collisions at RHIC and the LHC (Ref.~\ref{Sect:Colliders}) have generated vast, high-precision datasets to study the Quark-Gluon Plasma and to quantify its properties. High-energy nuclear collisions likewise probe the partonic structure of hadrons, and even provide complementary information to low-energy nuclear structure physics. However, collider experimental data are not directly interpretable in terms of physical properties of QCD many-body systems. Rather, theoretical modeling using Monte Carlo simulations is usually required to connect experimental measurements with QCD theory and QCD-based phenomenology. 

In order to be physically meaningful, these models are based on complex, detailed event generators, comprising a set of effective models in a unified numerical framework. Such simulations are computationally expensive, requiring exascale resources at HPC facilities to obtain the required precision. It is however challenging for nuclear physicists to develop and continually update the numerical simulation frameworks for effective utilization of continually evolving HPC facility hardware and infrastructure. 

Multidisciplinary collaborations, comprising domain scientists, computer scientists, and data scientists, have been formed to face this challenge. Long-term dedicated support for this approach is needed, to enable the field to fully exploit the large investment it has made in heavy-ion collider physics.

\subsection{Advanced computing in NP: experiment}
\label{Sect:ACexperiement}

While NP experimental projects receive significant funding from the DOE and NSF to support their core activities, they also require support for R\&D for advanced computing that is not covered by the core funding. This section sketches the advanced computing landscape in the NP experimental research areas.

\subsubsection{FRIB}
\label{Sect:FRIBcurrent}

FRIB (Facility for Rare Isotope Beams), on the campus of Michigan State University, is the newest scientific user facility of the DOE SC Office of Nuclear Physics, with more than 1,800 registered users (FRIBUO~\cite{FRIBUO}). FRIB enables scientists to make discoveries in the properties of atomic nuclei, in nuclear astrophysics, and in the fundamental symmetries of nature, using previously unavailable beams of rare isotopes. The FRIB scientific mission is a central element of the 2023 NSAC Long Range Plan for Nuclear Science~\cite{NSACLRP}.

Experiments with rare isotope beams are comparably short, with a typical duration of a few days to a week, and there are substantial benefits to developing and implementing novel accelerator control software based on artificial intelligence and machine learning (AI/ML) techniques to shorten machine setup time while maintaining high-quality, high-power accelerated ion beams. Specifically, AI/ML tools can address three critical areas: expediting accelerator switching between ion species, minimizing losses of multiple charge-state beams, and ensuring uninterrupted high-power beam delivery for physics experiments, ensuring a safe operation.  
 
Bayesian Optimization (BO) techniques for automated and efficient beam tuning include pyBO, a custom Python package that has been implemented in the FRIB EPICS control system.
Virtual diagnostics (VD) tools based on AI/ML algorithms have also been developed for non-invasive measurements of beam phase space parameters along the accelerator. In collaboration with DOE national laboratories, FRIB is developing conditional diffusion models for phase-space prediction generated by the Electron Cyclotron Resonance (ECR) ion source.
The ML-based VD techniques are crucial for the online running of the accelerator digital twin. 
Implementing and using AI/ML at FRIB requires a highly skilled workforce. Several staff physicists, postdocs, and DOE-supported Accelerator Science and Engineering Traineeship (ASET) graduate students are involved in developing AI/ML tools for beam tuning and accelerator operations.
 
FRIB’s science program also requires state-of-the-art scientific instrumentation aligned with the identified science drivers and increasing facility capabilities.
In addition to the state-of-the-art, day-one capabilities, new scientific instrumentation is continually being developed and constructed to harness the full discovery potential of the facility. During a typical experiment, datasets ranging in size between a few GB to ~100 TB are recorded, depending on the instrument, trigger, and science case.  Most FRIB experimental collaborations do not have dedicated software developers, and each experiment develops a tailored analysis pipeline based on its expected needs based on a Laboratory-supported data acquisition and analysis framework. The framework is developed and supported by a Scientific Software Group which also carries out R\&D for new data acquisition systems and analysis pipelines in response to evolving user requirements.
 
For several current FRIB analyses of large-volume data sets, investigators are applying AI/ML techniques to search for rare signals. Salient examples include the Active Target Time Projection Chamber (AT-TPC) and the FRIB Decay Station Initiator science programs.  The application of AI/ML approaches has the potential to significantly speed data analysis by orders of magnitude, enabling analyses that would otherwise not be feasible. Further development of generalized AI/ML frameworks is being pursued and will open new avenues to adapt AI/ML techniques across the evolving FRIB experimental program.

\subsubsection{JLab}
\label{Sect:JLabcurrent}

The Continuous Electron Beam Accelerator Facility (CEBAF) at Jefferson Lab (JLab)\cite{JLAB} is a unique facility for detailed exploration of the structure of hadrons and atomic nuclei, and other aspects of QCD and the Standard Model. CEBAF hosts about 1900 users from around the world. JLab also partners with BNL for the management, design and construction of the Electron Ion Collider.

CEBAF delivers a highly polarized electron beam, with currents up to 180 $\mu$A, to fixed targets in four experimental halls simultaneously.  Beam characteristics (intensity, polarization, and energy) are tuned to optimize the physics utilization of each hall independently. Data rates from the experimental halls have risen continuously as a result of new detector and data acquisition (DAQ) capabilities, and are projected to reach 3--4\,GB/sec in 2025.

Most of the data produced by JLab are stored and analyzed on-site. Currently, JLab's experimental program generates datasets on the order of 10–20 PB per year. The computing infrastructure for analysis and simulation is primarily housed within the JLab Scientific Computing facility. However, there is significant and increasing use of external resources such as the Open Science Grid (OSG) and the National Energy Research Scientific Computing Center (NERSC).  Software support includes a mix of internally supported common infrastructure (DAQ, data storage management, compute workflow management, etc.) and application-specific analysis and simulation toolkits managed by the external collaborations, often with Hall support.  This model works well for large collaborations, but presents an ongoing challenge for small collaborations with high personnel turnover rates.

JLab has made a significant investment in developing and delivering AI/ML supported tools, both internally and in cooperation with external collaborations.  These ongoing R\&D efforts are critical to take full advantage of the vast amount of data that modern detector systems provide, improving operational efficiency and reducing the time to final physics publication.

The High Performance Data Facility (HPDF)\cite{HPDF} launched in late 2023 will be hosted at JLab.  Once operational, this unique scientific user facility will provide advanced infrastructure for data-intensive science.

\subsubsection{Neutrino experiments}
\label{Sect:Neutrinoscurrent}

The NP neutrino physics portfolio encompasses neutrinoless double--$\beta$ decay $0\nu\beta\beta$ and sterile neutrino searches, and neutrino mass and scattering cross-section measurements. It has significant overlap with the HEP neutrino and dark matter programs, and with other Fundamental Symmetry efforts.

Compared to other NP sub-areas, experimental neutrino physics has a stronger emphasis on hardware and highly-customized detectors, with less emphasis on advanced computing and physics modeling.  Experimentally, events are rare and require minimal on-detector data reduction.  However, large-scale simulations are vital to understanding systematics and backgrounds. The models in these simulations are complex and highly tuned, and such calculations are computationally expensive. Next-generation $0\nu\beta\beta$ search detectors project data volumes of $\sim100$ TB/year (nEXO~\cite{nEXO}) or 1 PB/year (LEGEND-1000~\cite{LEGEND}).

Ongoing challenges in this sector include ensuring that the data sets are portable and well documented, allowing broader analysis access outside of a custom computing environment. Integration of AI/ML techniques for low-level data processing, exploration of systematics, and accelerated reconstruction shows significant promise.

\subsubsection{Collider experiments}
\label{Sect:Colliders}

Collider experiments at RHIC~\cite{RHIC} and the LHC~\cite{LHC}, and in future at the EIC~\cite{EIC}, are based on large multi-purpose detectors which run many months per year, managed by collaborations whose international membership numbers in the hundreds to few thousands. The data pipeline of modern detectors is typically continuous-streaming readout, with multiple levels of on-the-fly data reduction, analysis, and event tagging. 

As an example, the ALICE detector at the LHC~\cite{ALICE:2023udb} currently generates 3.4 TB/s from its readout electronics, which is reduced to less than 100 GB/s for transmission to long-term storage, corresponding to an annual recorded data volume of about 60 PB. ALICE reconstruction and analysis software consists of about 2M lines of code, managed by a core computing team of about ten people, with expert contributions from hundreds of collaborating physicists. The time interval from the recording of raw data to its availability for physics analysis is currently about one year, limited by the complexity of the data and by computing capacity in terms of CPU, data management, and storage. Computing capacity for simulations that are essential for data analysis are similarly limiting. The STAR and sPHENIX experiments at RHIC, the ATLAS, CMS, and LHCb experiments at the LHC, and the ePIC experiment at the future EIC, have similarly large data volumes, work flows, and computing challenges, though differing in detail in each case.

A collider experiment is a decades-long enterprise, requiring sustained support for computing infrastructure. This need persists after the completion of experimental running, for data preservation and re-analysis. Support for this effort is provided primarily by the host laboratory (BNL, CERN, JLab, etc.), which is also responsible for the retention of institutional memory and specific expertise via staffing. However, other institutions and funding agencies, beyond those supporting the host laboratory directly, typically also provide significant support for long-term software sustainability and distributed computing.

A common theme for software infrastructure in large international collaborations is the interplay between commercially-provided software solutions, publicly available open source software (e.g. GNU/GPL, CERN), and purpose-built solutions. The relative cost of the various approaches, in terms both of personnel effort and financial support, varies with time. Collaborations continuously evaluate such costs, and invest in software development to ensure long-term software functionality and to renew expertise within the collaborations. Essential software development includes both experiment-specific applications and innovative technologies, notably AI/ML and new computing hardware paradigms.


Recent efforts, such as those in the ePIC collaboration at the upcoming EIC, illustrate an emerging trend toward adopting sustainable software practices from the outset. By leveraging mature, actively maintained software stacks, such as those developed within the particle physics and broader data science communities, ePIC aims to reduce long-term technical debt and enhance maintainability. These choices are informed by a conscious effort to integrate community-supported tools rather than developing bespoke solutions unnecessarily, enabling more efficient development and distributed support through active engagement with external projects

\subsection{Advanced computing in NP: tools}
\label{Sect:ACtools}

\subsubsection{Bayesian Inference}

A broad range of problems in NP require the determination of  phenomenological constraints on multidimensional model parameters from complex experimental measurements, generically labeled ``inverse problems.''  One such example is the reconstruction of the internal quark and gluon structure of nucleons and nuclei using momentum--space and position--space tomography, based on measurements of quantum correlation functions (QCFs). A second example is the reconstruction of properties of the Quark-Gluon Plasma (QGP) based on collider data from nuclear collisions; the observational data are not directly sensitive to the physical properties of interest, and simulation-based inference is needed to reconstruct these quantities.

Bayes’s Theorem is a powerful tool to solve Inverse Problems, providing conceptually transparent and unbiased constraints on theoretical parameters and their uncertainties (“Bayesian Inference”), and enabling the quantification of agreement or tension between models and data. However, the application of Bayesian Inference to NP problems faces several challenges. The theoretical model may be computationally expensive; efficient exploration of the model parameter space may require ML/AI techniques, for instance the incorporation of training data at different fidelity levels. Alternatively, the model parameter dimensionality may be several orders of magnitude, rendering the convergence to a meaningful posterior distribution computationally challenging; ML/AI tools for analyzing high-dimensional probability distributions could reveal novel physics relations embedded in complex models. 

For hadronic structure, Bayesian inference computations are commonly rendered tractable by transforming observational data to a simplified space in which the NP theory calculations can be carried out directly, for extraction of distributions such as collinear Parton Distribution Functions. Recently, new data and theoretical developments have extended the scope of QCD global analysis, enabling the extract multi--dimensional partonic information. This enhanced scope is computationally demanding, and current efforts are limited by available computation capacity (Sect.~\ref{sketchJAM}).

For QGP studies, Bayesian Inference corresponds to calculation of the likelihood in a model parameter space of dimension $\sim5$ to 25, depending upon the problem being investigated. The computational challenge arises because a computationally--expensive physics simulation must be evaluated for each choice of parameter space--point. Efficient utilization of current HPC facilities, to achieve sufficient statistical precision for physically meaningful calculations, requires the application of AI/ML approaches (Sect.~\ref{sketchJETSCAPE}).

\subsubsection{Event generators}

Theoretical calculations of physical processes which are of interest in NP research are often not amenable to direct comparison with experimental measurements. For example, high-energy QCD processes such as jet production are measured by experiments at RHIC and the LHC as distributions of physical particles (hadrons), while the underlying theoretical calculations describe distributions of quarks and gluons. Connecting the two requires phenomenological modeling of hadronization, the process by which quarks and gluons are transformed into physically observable hadrons, which can only be carried out numerically.

Monte Carlo (MC) event generators are computer codes which encapsulate the physical picture underlying a certain class of processes, and generate the resulting particle distributions that would be seen by experiment. Event generators are important in several NP sub-areas for bridging theory and experiment, for exploring the consequences of modeling assumptions and model parameter dependencies, and for analyzing experimental data. Event generators play a vital role in collider~\cite{PYTHIA,HERWIG,Campbell:2022qmc,Deng:2010mv,JETSCAPE} and neutrino physics~\cite{Andreopoulos:2009rq} studies. There is at present a need for extensive development of MC generators for the future Electron–-Ion Collider (see e.g.~\cite{MCforEIC}). 

Event generator modeling of a full high--energy collision, from the structure of the projectiles through final state particle distributions, is complex. Heavy–ion event generators simulate multiple distinct nuclear collision stages, including projectile parton distributions, partonic scatterings, thermalization of the resulting partonic gas, QGP formation and collective dynamics, jet-QGP interactions, hadronization, and hadronic decay (e.g.~\cite{Wang:1991hta,Pierog:2013ria,Bierlich:2018xfw}). Generators for ultra-peripheral collisions must account for photon polarization, and the quantum mechanical interference between the two production sites \cite{Klein:2016yzr,Burmasov:2021phy,Shao:2022cly}. 

Generators for $ep/eA$ at the EIC must include interactions at various energy scales corresponding to hard photon-parton interaction, hadronization, and nuclear breakup \cite{Lomnitz:2018juf,Chang:2022hkt,Aschenauer:2022aeb}. For the study of gluon saturation and non-linear QCD evolution, which will be explored at the EIC and by forward measurements at the LHC, event generators must incorporate theoretical calculations based on the BK dipole evolution equation and its generalization, JIMWLK (Sect.~\ref{Sect:partonstructure}). These calculations entail high-order numerical integrations which require ML techniques for efficient and accurate solution on current HPC facilities.

The development and maintenance of a complete event generator requires a dedicated group effort, and such efforts are relatively few in number. General--purpose event generators serve a large and diverse user community, which requires support for multiple operating systems and compilers, and continuous updates as operating systems, compilers, and hardware evolve. These codes must be maintained over a long time period; the most popular Monte Carlos have been in usage for over 30 years.  

Despite the importance of MC event generators to the NP program and their large user community, support for the writing and stewardship  of MC event generators has traditionally been challenging to obtain.

\subsubsection{Quantum information science}
             
Nuclear physicists are at the forefront of advancing quantum information science, engineering and technology, through their close collaborations among technology companies, national laboratories and universities. These collaborations have significantly progressed basic research in nuclear physics, for example in simulating the dynamics of nuclei, quarks and gluons, and neutrinos, relevant for the structure of matter and for transport in extreme conditions. This has been facilitated by DOE NP, DOE ASCR, NQI Centers, DOE Quantum Testbed programs, and by premier or direct access to forefront quantum computers such as cold-atom systems, superconducting radio frequency (SRF)-cavity systems, trapped-ions, superconducting-qubit systems. These accomplishments have accelerated similar efforts in other domain sciences and engineering, and in quantum sensing and communication. 

Error mitigation techniques used in present-day NISQ-era simulations are based on algorithms developed by scientists at technology companies (e.g., IBM) and by nuclear and high-energy physicists for specialized applications, while error-correction protocols under development for logical qubits and operations require further developments for specialized applications in NP, including for Gauss’s law checks in LGTs and hierarchical implementations in low-energy systems. The hardware and software configurations which are currently available for practical use are somewhat patchwork, and improved organization is needed for significant future progress, particularly for the near-term transition from NISQ-era to fault-tolerant/error corrected quantum computers.

The evolution from noisy, intermediate scale, quantum NISQ-era quantum computers to fault-tolerant/error-corrected quantum computers is leading nuclear physicists to now focus on customized, physics-aware information encoding, communication, and error-correction, to address these challenges.

\subsection{Survey of term-funded projects}
\label{Sect:FundedProjects}

Advanced Computing in modern NP research covers a broad scientific scope. Projects that are formally funded by the instruments outlined in Sect.~\ref{Sect:FundingInstruments} play a special role in advancing the frontiers of advanced NP computing. The workshop devoted significant plenary time to surveying currently funded projects, with emphasis on their computing scope and future needs.

This section presents brief sketches of the projects discussed at the workshop. While not a complete list of all US-funded NP advanced computing projects, it provides a representative cross-section. The interested reader should consult the workshop presentation slides~\cite{SANPCindico} and project web pages for further detail.

\subsubsection{BAND}
\label{sketchBAND}

The BAND Collaboration~\cite{BAND,BANDmanifesto} (Bayesian Analysis of Nuclear Dynamics) seeks to create a suite of general-purpose software tools that will enable principled Uncertainty Quantification across a wide range of NP problems, including neutrinoless double beta decay;  r-process nucleosynthesis and the associated nuclear masses, and reaction and beta-decay rates; heavy-ion collisions and inference of Quark-Gluon Plasma properties; as well as reaction dynamics and nuclear data evaluation. Significant emphasis is placed on code sustainability, code stewardship, and the reproducibility of results. Many of the problems considered require quantified extrapolation.

BAND is funded by the NSF CSSI program. BAND currently has 32 collaborators at 11 institutions.

\subsubsection{ENAF}
\label{sketchENAF}

The ENAF Collaboration~\cite{ENAF} (Exascale Nuclear Physics for FRIB) studies the production of chemical elements in astrophysical events using cutting-edge astrophysics simulations that employ state-of-the-art nuclear physics. 
The primary focus of ENAF is the r-process in neutron star  mergers. ENAF also seeks to understand the impact of neutrino flavor oscillations on this nucleosynthesis mechanism, as well as its observability in kilonovae.  
Technical challenges include high-performance GPU-based implementations of key computational kernels for thermonuclear kinetics and neutrino radiation transport computations, and ML-based inference engines designed for the interpolation of tabular astrophysics data. 
The maintenance of high performance on evolving HPC architectures is challenging for such multi-physics codes. 

ENAF is supported by DOE/NP and ASCR through the SciDAC program. ENAF currently has 14 collaborators at 6 institutions. 

\subsubsection{IQuS}
\label{sketchiQuS}

The InQubator for Quantum Simulation (IQuS)~\cite{IQUS} at the University of Washington carries out theoretical studies of strongly-interacting correlated matter and complex quantum systems, using quantum simulations and theoretical approaches that incorporate quantum entanglement and coherence. The IQuS scientific scope includes lattice gauge theories, nuclear structure and reactions, spin systems, cold atoms, neutrinos, topological phases and dynamics, quantum complexity, and hardware development. The institute currently has seven staff and three postdocs, with a workshop and visitor program carried out in conjunction with the Institute for Nuclear Theory at UW. 

Continued NP support is needed for access to forefront and developmental quantum computer architectures, and to high-performance computing and software. Such support would enable  researchers to address challenges which are central to NP research, including non-equilibrium dynamics of dense systems of neutrinos, low-energy nuclear reactions important for advanced nuclear reactor design, and the dynamics of the Quark--Gluon Plasma.

\subsubsection{JAM}
\label{sketchJAM}

The JAM Collaboration~\cite{JAM} (Jefferson Lab Angular Momentum) studies the internal quark and gluon structure of hadrons by reconstructing quantum correlation functions (PDFs, FFs, TMDs, GPDs) from experiment. This inference task, corresponding to the solution of multiple layers of an inverse problem (factorization, evolution, $\ldots$), is challenging. 

JAM takes a rigorous, computationally intensive Bayesian Monte Carlo approach. JAM calculations have reached the limit of computing resources available from the JLab CST Division. ML-based techniques are being explored to address this limitation, including collaboration with other projects (QGT, JLab LDRD, QuantOm) to optimize GPU hardware usage. Additional synergistic collaboration between NP and ASCR-supported projects is needed to fully meet the promise of the JLab 12 GeV program

The JAM collaboration includes theorists, experimentalists, and computer scientists, with 27 collaborators at 13 institutions. The collaboration is hosted at Jefferson Lab, which coordinates development of software and provides access to its HPC systems. The JAM Collaboration has been funded by DOE and NSF. While the collaboration funding support period has ended, the Collaboration continues to work jointly on common projects.

\subsubsection{JETSCAPE}
\label{sketchJETSCAPE}

The JETSCAPE Collaboration~\cite{JETSCAPE,JETSCAPEpubs} 
(Jet Energy-loss Tomography with a Statistically and Computationally Advanced Program Envelope) studies hot and cold QCD matter using a modular framework for simulating the complex dynamical environment of high-energy hadronic and nuclear collisions, and Bayesian inference for rigorous data/model comparisons. JETSCAPE focused initially on hot QCD matter and the Quark-Gluon Plasma, but more recently has expanded its scope to include the study of both QCD phase transitions and the partonic structure of nucleons and nuclei. 

The Bayesian inference component of the JETSCAPE scope is computationally expensive,  and is limited by the availability of HPC resources. JETSCAPE and related efforts develop and exploit ML-based approaches for emulation and likelihood evaluation, to achieve efficient resource utilization. 


The JETSCAPE Collaboration is an interdisciplinary team of theoretical and experimental physicists, computer scientists, and statisticians, comprising about 60 collaborators from 13 institutions. JETSCAPE has been funded by the NSF/CSSI program. Continued support is vital to maintain and develop the large and complex JETSCAPE software framework, thereby capitalizing on this investment.

\subsubsection{MUSES}
\label{sketchMUSES}

The MUSES Collaboration~\cite{MUSES} (Modular Unified Solver of the Equation of State) develops cyber-infrastructure to address interdisciplinary questions in nuclear physics, gravitational wave astrophysics, and heavy-ion physics, with special focus on the QCD phase diagram and neutron stars, using tools provided by lattice and perturbative QCD, chiral effective field theory and models \cite{MUSES:2023hyz}. MUSES creates software communities through short--term grants to academic groups for software development related to their research. Such efforts in part build upon existing software ecosystems, but also create new ones. The resulting software is free and open--source.

There are three core aspects to the MUSES cyberinfrastructure: independent software modules that calculate equations of state, calculate physical observables, and/or process data; a software framework for integration and interoperability of these modules; and the Calculation Engine, a management system to orchestrate the execution of composable, multi-module workflows.


MUSES is funded by the NSF/CSSI program. MUSES is an interdisciplinary team that currently counts 134 collaborators, 25\% of whom carry out software development, from about 20 institutions. 

\subsubsection{NUCLEI}
\label{sketchNUCLEI}

The NUCLEI Collaboration~\cite{NUCLEI} (Nuclear Computational Low-Energy Initiative) carries out precise calculations, with quantified uncertainties, of stable nuclei and rare isotopes, the electroweak response of nuclei and nuclear matter, and nuclear fission. The collaboration develops and refines low-energy nuclear interactions and currents, exploiting exascale computing facilities for the modeling of atomic nuclei, their reactions and decay, and nuclear matter.  These calculations support experimental programs at FRIB, ATLAS, JLab, DUNE, and future ton-scale neutrino facilities. 

NUCLEI is funded by the SciDAC program, with active participants from both RAPIDS and FASTMath. It is an interdisciplinary collaboration of physicists, computer scientists, and applied mathematicians, with 57 members at 13 institutions. 

\subsubsection{QuantOm}
\label{sketchQUANTOM}

The QuantOm Collaboration~\cite{QUANTOM} (QUAntum chromodynamics Nuclear TOMography) develops novel approaches to the femtoscale imaging of protons and nuclei in terms of their three-dimensional quark and gluon structure. Key questions include the momentum and spatial distribution of quarks and gluons, and the dynamical generation of proton mass. 

QuantOm focuses on optimizing the comparison of theoretical calculations and experimental electron-scattering measurements. Current NP and HEP published experimental data are reported as binned histograms, resulting in a significant loss of information and analysis flexibility. QuantOm is developing mechanisms for unbinned, event-wise analysis of large datasets from JLab and the future EIC, to overcome these limitations. This approach is computationally intensive, requiring exascale resources and new ML/AI tools for data storage and analysis.

QuantOm is supported by the SciDAC program. QuantOm is an interdisciplinary collaboration of QCD theorists and experimentalists, applied mathematicians, data scientists, and HPC experts. The QuantOm collaboration has 25 members from 6 institutions.

\subsubsection{USQCD}
\label{Sect:LQCDintro}


Current Lattice QCD (LQCD) calculations cover several NP sub-areas, with strong connections to the NP and HEP experimental programs~\cite{LatticeQCD,Beane:2010em,Detmold:2019ghl, Constantinou:2022yye, USQCD:2022mmc}:

\begin{itemize}
\item QCD phase structure and thermodynamics: confinement and chiral symmetry phase transitions; QCD critical point; color superconductivity; neutron star equation of state; transport in the Quark-Gluon Plasma. Synergy with the heavy-ion experimental programs at RHIC and the LHC.
\item Hadron spectroscopy: hadron masses; prediction of new heavy or exotic hadrons. Synergy with the JLab and EIC experimental programs.
\item Hadron structure: EM and Gravitational Form Factors; partonic structure (PDFs, TMDs, GPDs). Synergy with the JLab and EIC experimental programs.
\item Fundamental Symmetries: lepton number violation; properties of the neutrino. Synergy with neutrinoless double beta decay ($0\nu\beta\beta$) experimental program. 
\item Beyond Standard Model (BSM) physics: neutrinos; dark matter; lepton flavor violation. Synergy with DUNE, HyperK, dark-matter searches, $\mu{2}e$.
\end{itemize}


The US LQCD community effort is coordinated by the USQCD Collaboration~\cite{USQCD,USQCDCharter}, whose membership of 180 scientists comprises over 90\% of all US LQCD researchers in NP and HEP. The USQCD Collaboration fosters software development and student and postdoc training, and coordinates the funding and allocation of mid-scale computing resources located at JLab, BNL, and FNAL. The US LQCD effort has been supported by SciDAC for over two decades, with contributions from NP, HEP, ASCR, and NNSA. 


The US LQCD community has significant connections to industry, notably GPU library development (QUDA/NVIDIA~\cite{QUDA}) and code generators for new hardware. The flow of LQCD expertise from academia to industry helps foster these connections.

\subsection{Workshop contributed talks}
\label{Sect:ContributedTalks}

The workshop included an open mic session, with brief contributed talks. This section summarizes key points raised in that session about future directions in advanced computing for NP.

\paragraph{Challenges in non-equilibrium many-body dynamics (A. Bulgac)} Many physical phenomena in NP involve non-equilibrium long-term dynamics, e.g. nuclear fission, Quark-Gluon Plasma evolution, quantum turbulence, many-body entanglement, fermion-antifermion production in strong fields, neutrino collective oscillations, fermionic superfluid systems, quantum transport, and non-Markovian processes. Theoretical approaches to such processes based on traditional calculational methods, which describe stationary or quasi-stationary processes, are inadequate, necessitating a qualitatively new approach.

Such calculations require HPC capability at the largest scale available, and will do so for the foreseeable future. These are long-term projects, requiring long-term code development and data storage. Funding instruments are needed to support and steward such efforts in a sustained way, including that of small research groups.

\paragraph{Machine Learning for NP applications (M. Ploskon)}
There are applications of ML that have the potential to be transformative in NP research, but that are not of similar importance to industry. Examples include explainable and interpretable ML (e.g.~\cite{Lai:2020byl}) and novel data analysis architectures using graph Neural Networks.

Interdisciplinary collaboration with computer scientists has proven to be effective in addressing such problems. However, such efforts correspond more to the application of existing ML approaches to NP problems, and less to the development of novel techniques. Current funding instruments available to NP domain scientists are on the whole not targeted at such ML applications, as opposed to more generic ML developments; nevertheless, ML applications are increasingly important to advance NP science, in both experiment and theory. A support stream for ML-related applications to advance NP science would be beneficial.

\paragraph{Open data for NP simulations (C. Shen)}
Bayesian inference analysis of heavy-ion collision data requires large-scale computing resources, generating large volumes of simulated data~\cite{JETSCAPE:2020mzn,JETSCAPE:2024cqe}. Such analyses employ ML-based techniques for efficient utilization of computing capacity, and open access for community usage of expensive simulated data such as these would be highly beneficial to the field.

However, data storage is expensive, and both general access and long-term data storage require robust software solutions.
As an example, there are several possible approaches to address this issue for Bayesian inference of heavy-ion data, listed here from simple to comprehensive:

\begin{itemize}
\item tagged version of code container and parameter data stored on the Open Science Grid. This is appropriate for model emulation in Bayesian Inference analyses.
\item Particle yields and flow vectors as functions of kinematic observables for each simulated event. This is appropriate for general model emulation at the event-by-event level and is an effective way to communicate about new experimental analyses. 
\item All final-state hadron momentum information for each event, with end-to-end event evolution history. This is appropriate only for the most detailed analyses, and only a small fraction of the data is needed in this format.
\end{itemize}

A support stream to develop and support open data access, both for heavy-ion physics research and for other sub-fields, would be beneficial.

\paragraph{Community Driven Repositories of Knowledge (K. Godbey)} The Advanced Scientific Computing and Statistics Network (ASCSN)~\cite{ASCSN} fosters collaborations and builds connections between developers in computer science, machine learning, statistics, and applied mathematics, and in application domains which benefit from such developments. The ASCSN scope also includes Cloud Computing and Machine Learning. 

The ASCSN community-driven forum has been used to support FRIB summer schools~\cite{FRIBsummerschool}. Pedagogical resources of this kind lower the entry barrier for scientists across a wide spectrum of institutions, backgrounds, and career stages. However, there is currently no established funding framework for the sustained support of such efforts. This arises in significant part because their scientific focus is cross-disciplinary, at the boundaries of nuclear physics, computer science, and data science, and therefore not clearly within the purview of any one funding agency. New funding approaches to address such cross-disciplinary pedagogical efforts would be beneficial.

\newpage
\appendix
\section{List of acronyms}
\label{sect:Acronyms}

\begin{itemize}

\item AI: Artificial intelligence
\item ASCR: Advanced Scientific Computing Research program
\item BNL: Brookhaven National Laboratory
\item BSM: Beyond Standard Model
\item CEBAF: Continuous Electron Beam Accelerator Facility
\item CERN: European Council for Nuclear Research
\item CI: CyberInfrastructure 
\item CSSI: Cyberinfrastructure for Sustained Scientific
Innovation
\item DAQ: Data acquisition
\item DESY: Deutsches Elektronen-Synchrotron (German Electron Synchrotron)
\item DNN: Office of Defense Nuclear Nonproliferation
\item EIC: Electron-Ion Collider
\item FNAL: Fermi National Laboratory
\item FRIB: Facility for Rare Isotope Beams
\item HEP: High Energy Physics
\item HERA: Hadron–Electron Ring Accelerator
\item HPC: High-Performance Computing
\item JLab: Jefferson Laboratory
\item LDRD: Laboratory Director's R\&{D} funds
\item LEP: Large Electron-Positron Collider
\item LHC: Large Hadron Collider
\item LRP: Long Range Plan
\item LQCD: Lattice QCD
\item ML: Machine Learning
\item NERSC: National Energy Research Supercomputing Center
\item NNSA: National Nuclear Security Administration
\item NP: Nuclear Physics
\item NSAC: Nuclear Science Advisory Committee
\item OSG: Open Science Grid
\item QCD: Quantum Chromo-Dynamics
\item QED: Quantum Electro-Dynamics
\item QGP: Quark-Gluon Plasma
\item RHIC: Relativistic Heavy-Ion Collider
\item SciDAC: Scientific Discovery through Advance Computing
\end{itemize}


\section{Workshop organization and agenda}

\subsection{Workshop organizing committee}
\begin{itemize}
\item Amber Boehnlein (JLab), co-chair
\item Peter Jacobs (LBNL), co-chair
\item Joe Carlson (LANL)
\item Ian Cloet (ANL)
\item Markus Diefenthaler (JLab)
\item Robert Edwards (JLab)
\item Raphael Hix (ORNL)
\item Thomas Papenbrock (UTK)
\item Brad Sawatzky (JLab)
\item Torre Wenaus (BNL)
\end{itemize}

\subsection{Workshop agenda}

Workshop indico page:~\cite{SANPCindico}.

Agenda: 

\begin{longtable}{%
        >{\raggedright}p{0.1\textwidth} 
        >{\raggedright}p{0.6\textwidth} 
        l}
\hline 
\multicolumn{3}{c}{\bf Plenary: Thursday, June 20} \\ 
{\bf Time} & {\bf Title} & {\bf Presenter}\\
\hline 
09:00 & Welcome     & SAWATZKY, Brad\\ 
09:05 & Introduction to SANPC & BOEHNLEIN, Amber \\ 
09:20 & MUSES & MANNING, T. Andrew \\ 
09:40 & ENAF & HIX, W Raphael \\ 
10:00 & NUCLEI & PAPENBROCK, Thomas \\ 
10:20 & \multicolumn{2}{c}{\emph{Break}} \\ 
10:50 & BAND & GODBEY, Kyle \\ 
11:10 & JETSCAPE & MAJUMDER, Abhijit \\ 
11:30 & QuantOm & CLOET, Ian \\ 
11:50 & SciDAC FASTMath & MUNSON, Todd\\ 
12:10 & Infrastructure for Nuclear Nonproliferation Data Science Research & ADAMSON, Paul \\ 
12:30 & \multicolumn{2}{c}{\emph{Lunch}} \\ 
13:30 & Exp - Collider (RHIC, LHC, EIC) & PLOSKON, Mateusz \\ 
13:55 & Exp - JLab/FRIB & LIDDICK, Sean\\ 
14:20 & Computing for Neutrino Physics & BRODSKY, Jason\\ 
14:55 & \multicolumn{2}{c}{\emph{Break}} \\ 
15:25 & JLab Angular Momentum (JAM) & MELNITCHOUK, Wally\\ 
15:45 & LQCD & DETMOLD, William\\ 
16:15 & iQUS & SAVAGE, Martin \\ 
16:35 & Community-driven repositories of knowledge & GODBEY, Kyle \\ 
16:45 & Sharing simulation data with community & SHEN, Chun\\ 
16:55 & Maintaining and Refactoring Legacy Code & FLORES, Abraham \\ 
17:05 & Challenges in non-equilibrium many-body dynamics & BULGAC, Aurel\\ 
17:15 & HPC and Quantum Computing in Nuclear Dynamics & CARLSON, Joseph\\ 
17:25 & QMC calculations of nuclear responses: beyond Carbon & ANDREOLI, Lorenzo\\ 
17:35 & ML for [high-energy] Nuclear Physics & PLOSKON, Mateusz\\ 
17:45 & Neural Network for low-energy nuclear structure & GNECH, Alex\\ 
17:55 & Optimization with Correlated Errors & SOLTZ, Ron \\ 
18:05 & TListSpectrum: toolset for spectroscopic analysis & GARLAND, Heather \\ 
18:15 & Future proofing your research & BROWN, David\\ 
\hline 
\end{longtable}

\begin{longtable}{%
        >{\raggedright}p{0.1\textwidth} 
        >{\raggedright}p{0.6\textwidth} 
        l}
\hline
\multicolumn{3}{c}{\bf Plenary: Friday, June 21} \\
{\bf Time} & {\bf Title} & {\bf Presenter}\\
\hline
08:30 & Software Stewardship at ASCR & DUBEY, Anshu \\
08:50 & Composable Optimization and Control Toolkit for Scientific Applications & RAJPUT, Kishansingh \\
09:10 & Data Science: Unique NP Challenges & NACHMAN, Ben \\
09:30 & Panel Discussion with Agencies & \\
\hline
\end{longtable}


\newpage
{\footnotesize 
\begin{longtable}{%
        >{\raggedright}p{0.1\textwidth} 
        >{\raggedright}p{0.5\textwidth} 
        l}
\hline
\multicolumn{3}{c}{\bf Parallel Sessions: Friday, June 21} \\
{\bf Time} & {\bf Title} & {\bf Organizers}\\
\hline
10:50 & Session 1: Experiment &  DIEFENTHALER, Markus \emph{and} \\
    && \hspace*{1cm}PLOSKON, Mateusz \\
10:50 & Session 2: Theory & RATTI, Claudia \emph{and} \\
    && \hspace*{1cm}ORGINOS, Kostas \\
13:30 & Session 1: Experiment &  DIEFENTHALER, Markus \emph{and} \\
    && \hspace*{1cm}PLOSKON, Mateusz \\
13:30 & Session 2: Theory & RATTI, Claudia \emph{and}  \\
    && \hspace*{1cm}ORGINOS, Kostas \\
13:30 & Session 3: Joint Theory and Experiment & GODBEY, Kyle \emph{and}  \\
    && \hspace*{1cm}SOLTZ, Ron \\
\hline
\end{longtable}
} 

\begin{longtable}{%
        >{\raggedright}p{0.1\textwidth} 
        >{\raggedright}p{0.6\textwidth} 
        l}
\hline
\multicolumn{3}{c}{\bf Plenary: Saturday, June 22} \\
{\bf Time} & {\bf Title} & \\
\hline
09:00 & White Paper open discussion & \\
12:00 & Adjourn & \\
\hline
\end{longtable}



\bibliographystyle{apsrev4-2}
\bibliography{SANPC}

\end{document}